\documentclass{Interspeech}
\usepackage{times}
\usepackage{latexsym}
\usepackage{booktabs}
\usepackage{verbatim}
\usepackage{multirow}
\usepackage[table,xcdraw]{xcolor}
\usepackage{colortbl}  
\usepackage{booktabs}
\usepackage{algorithm}
\usepackage{algpseudocode}

\usepackage{etoolbox}
\makeatletter
\patchcmd{\affiliation}
  {\renewcommand{\affiliationsep}{\vskip1pt}}
  {\renewcommand{\affiliationsep}{, }}
  {}{}
\makeatother

\usepackage[T1]{fontenc}
\usepackage[utf8]{inputenc}
\usepackage{microtype}
\usepackage{xcolor}
\definecolor{mustard5}{RGB}{255,250,205}    % Lightest (5%)
\definecolor{mustard10}{RGB}{255,240,150}    % Intermediate (10%)
\definecolor{mustard15}{RGB}{255,220,100}    % Darkest (15%)
\definecolor{mustardPP5}{RGB}{255,250,205}    % Lightest (5%)
\definecolor{mustardPP10}{RGB}{255,240,150}   % 10%
\definecolor{mustardPP15}{RGB}{255,220,100}   % 15%
\definecolor{mustardPP20}{RGB}{255,200,50}    % Darkest (20%)
\usepackage{inconsolata}
\usepackage{adjustbox}
\usepackage{subcaption, url, float, etoolbox, adjustbox, pgf,booktabs, verbatim}
\usepackage{xcolor}
\usepackage{amsmath}
\usepackage{amssymb}
\usepackage{tcolorbox}

\title{Towards Machine Unlearning for Paralinguistic Speech Processing}
\author[affiliation={1}]{Orchid Chetia}{Phukan*}
\author[affiliation={1,2}]{Girish*}{}
\author[affiliation={1,3}]{Mohd Mujtaba}{Akhtar*}
\author[affiliation={4}]{Shubham}{Singh}
\author[affiliation={5}]{Swarup Ranjan}{Behera}
\author[affiliation={6}]{Vandana}{Rajan}
\author[affiliation={7}]{Muskaan}{Singh}
\author[affiliation={1}]{Arun Balaji}{Buduru}
\author[affiliation={8,9}]{Rajesh}{Sharma}
\affiliation{}{IIIT-Delhi}{India}
\affiliation{}{UPES}{India}
\affiliation{}{V.B.S.P.U}{India}
\affiliation{}{BIT Mesra}{India}
\affiliation{}{Independent Researcher}{India}
\affiliation{}{Independent Researcher}{UK}
\affiliation{}{Ulster University}{UK}
\affiliation{}{University of Tartu}{Estonia}
\affiliation{}{Plaksha University}{India}
\email{\textcolor{blue}{\texttt{Correspondence:}} orchidp@iiitd.ac.in}

\usepackage{comment}

\begin{document}

\maketitle
\maketitle
\begingroup
  % switch to symbol‐style so \footnotetext uses “*” for the first footnote
  \renewcommand{\thefootnote}{\fnsymbol{footnote}}
  \setcounter{footnote}{0}
   \footnotetext{* Contributed equally as a first authors.}
\endgroup
\begin{abstract}
\noindent In this work, we pioneer the study of Machine Unlearning (MU) for Paralinguistic Speech Processing (PSP). We focus on two key PSP tasks: Speech Emotion Recognition (SER) and Depression Detection (DD). To this end, we propose, \textbf{\texttt{SISA++}}, a novel extension to previous state-of-the-art (SOTA) MU method, SISA by merging models trained on different shards with weight-averaging. With such modifications, we show that \textbf{\texttt{SISA++}} preserves performance more in comparison to SISA after unlearning in benchmark SER (CREMA-D) and DD (E-DAIC) datasets. Also, to guide future research for easier adoption of MU for PSP, we present \textit{``cookbook recipes''} - actionable recommendations for selecting optimal feature representations and downstream architectures that can mitigate performance degradation after the unlearning process. 
\end{abstract}

\section{Introduction}

\noindent The widespread use of ML models in applications ranging from personalized recommendations to health diagnostics has brought privacy issues to the forefront. As these models rely on vast amounts of personal data to optimize their performance, ensuring the ethical handling of such data has become a critical challenge. With such comes the risk of leakage of personal data, which can lead to identity theft, financial fraud, or other malicious exploitation. Furthermore, there has been a significant rise in adversarial attacks on ML models, specifically, membership inference attacks (MIA) \cite{shokri2017membership} where attackers makes efforts for extracting sensitive information from the models. These attacks not only compromise model integrity but also pose a serious threat to user privacy and trust in AI systems, highlighting the urgent need for enhanced model robustness and security measures. \par

In response, various privacy-preserving techniques such as homomorphic encryption \cite{mohammadi2023secure}, federated learning \cite{tsouvalas2022privacy}, and differential privacy \cite{dwork2008differential} have been developed to mitigate these risks. However, the balance between maintaining high-performance and ensuring robust privacy protection remains difficult to achieve \cite{mohammadi2023secure}. Also, one notable issue in this context is the is the principle of the \textit{right to be forgotten} bestowed to individuals within regulations like the GDPR \cite{mantelero2013eu}. This regulation empowers individuals to request the removal of their personal data from training datasets. However, retraining a model to comply with such a request is computationally expensive and unfeasible when datasets grow to large sizes and use of larger models for better performance. MU, by enabling targeted removal of specific data points, offers an elegant solution to this problem \cite{sommer2020towards, bourtoule2021machine, nguyen2022survey, lin2023erm, li2024machine}. MU is primarily categorized into three types: model-agnostic, model-intrinsic, and data-driven methods. Model-agnostic approaches \cite{koh2017understanding}, enable selective data removal across architectures. Model-intrinsic methods \cite{liu2024backdoor}, modify internal structures for unlearning. Data-driven techniques \cite{chen2024fast}, focus on data partitioning or augmentation.  While MU has been widely explored in domains like recommendation systems \cite{li2024making}, image classification \cite{lin2023erm}, and NLP \cite{wang-etal-2023-kga} %liu2023muter, yao2024machine},
, its application to speech-based tasks, particularly in paralinguistics, has not been fully explored. Paralinguistic speech processing (PSP) centers around the retrieval of information outside of literal speech content and has wide range of applications ranging from entertainment industry to healthcare. In this study, we introduce Machine Unlearning (MU) to the field of PSP for the first time. PSP encompasses tasks such as Speech Emotion Recognition (SER) and Depression Detection (DD), which rely on prosodic features of speech and present unique ethical challenges. The intimate nature of voice data, which carries sensitive emotional and mental health information, makes it particularly susceptible to unintentional memorization by machine learning models. This raises significant privacy concerns if such data is misused or inappropriately retained. Additionally, ML models built for PSP tasks face growing vulnerabilities to adversarial attacks \cite{alsenani23_interspeech, alsenani2024assessing}. Given these challenges, application of MU to PSP tasks offers a promising solution. As this approach allows models to selectively forget sensitive voice samples while maintaining their overall performance capabilities \cite{scherer2015self, gooding2021ethics}. \par
We focus on two critical PSP tasks: SER and DD. Additionally especially given the current research trend of leveraging large pre-trained models (PTMs) as feature extractors for enhanced performance benefits, MU is more beneficial as retraining models with these PTMs is cost-inefficient. This calls for unlearning novel MU methods that provides efficient retraining for unlearning as well as retains the performance after the unlearning process. To our end, we present, \texttt{\textbf{SISA++}}, a novel extension of the previous state-of-the-art (SOTA) MU method, SISA \cite{bourtoule2021machine}. \texttt{\textbf{SISA++}} introduces a critical enhancement by employing weight averaging to merge models trained on separate data shards\footnote{shard or subset term is used interchangebly}, effectively consolidating knowledge while maintaining modularity. This approach reduces inconsistencies and preserves the performance of the overall model after data unlearning. With these refinements, \texttt{\textbf{SISA++}} achieves superior performance retention compared to SISA as evidenced by evaluations on benchmark datasets such as CREMA-D (SER) and E-DAIC (DD). Furthermore, to facilitate widespread adoption of MU in PSP applications, we provide insights in the form of \textit{``cookbook recipes''}. These include step-by-step recommendations for selecting suitable feature representations, designing effective downstream architectures, and mitigating potential performance degradation caused by unlearning processes. These additions ensures that our study also serves as a resourceful guide for future researchers. 

\noindent \textbf{The key contributions of this paper are:}
    \begin{itemize}
    \item We pioneer the application of MU in PSP, specifically focusing on SER and DD.

    \item  We propose, \texttt{\textbf{SISA++}}, an MU method utilizing weight averaging to merge models trained on data shards, demonstrating superior performance retention compared to SISA on benchmark datasets (CREMA-D and E-DAIC).

    \item We provide \textit{``cookbook recipes''} - actionable guideline swith practical recommendations for selecting optimal feature representations and downstream architectures, enabling easier adoption of MU while mitigating performance degradation. 

    \item We give a comprehensive comparative investigation of features from various PTMs and downstream networks to find out the most optimal pairs as part of \textit{“cookbook recipes”}. TRILLsson features with transformer as downstream network are the most robust to the unlearning process. 
\end{itemize}
\noindent The code and models developed in this work are publicly available at: \url{https://github.com/Helix-IIIT-Delhi/SISA-Unlearning}

% To further promote transparency and reproducibility in the field, we will open-source the models and code used in this research study. This will allow researchers to build upon our work.

\section{Methodology}

In this section, we discuss preliminaries on SISA, the proposed novel extension, \textbf{\texttt{SISA++}} and lastly followed by the \textit{``cookbook recipes''}.

\subsection{Preliminary on SISA}
SISA \cite{bourtoule2021machine} is a SOTA data-driven technique for MU. It employs a data partitioning strategy with structured model training to enable efficient unlearning. The initial dataset \( D = \{(x_i, y_i)\}_{i=1}^{N} \), where \( x_i \) represents input features and \( y_i \) the corresponding labels, is split into \( K \) disjoint shards, denoted as \( D_k = \{(x_{ik}, y_{ik})\}_{i=1}^{N_k} \), ensuring that each shard contains a subset of the data without replacement. Independent sub-models \( \mathcal{M}_i \) are trained on these shards without communication between them. This structured approach facilitates efficient unlearning through partition-based retraining, requiring only the sub-models corresponding to the impacted shards to be updated when a data point \( (x_i, y_i) \) is removed or modified. The updated dataset \( D^{-i} \) is used to refresh the necessary sub-models while preserving the others. By minimizing the need for full model retraining, SISA significantly enhances computational efficiency and scalability, making the unlearning process more resource-efficient. During inference, a new input \( x \) is passed through all sub-models, and their outputs are aggregated using majority voting for classification and averaging for regression. 

\begin{figure}[!hbt]
    \centering
    \includegraphics[width=0.6\linewidth]{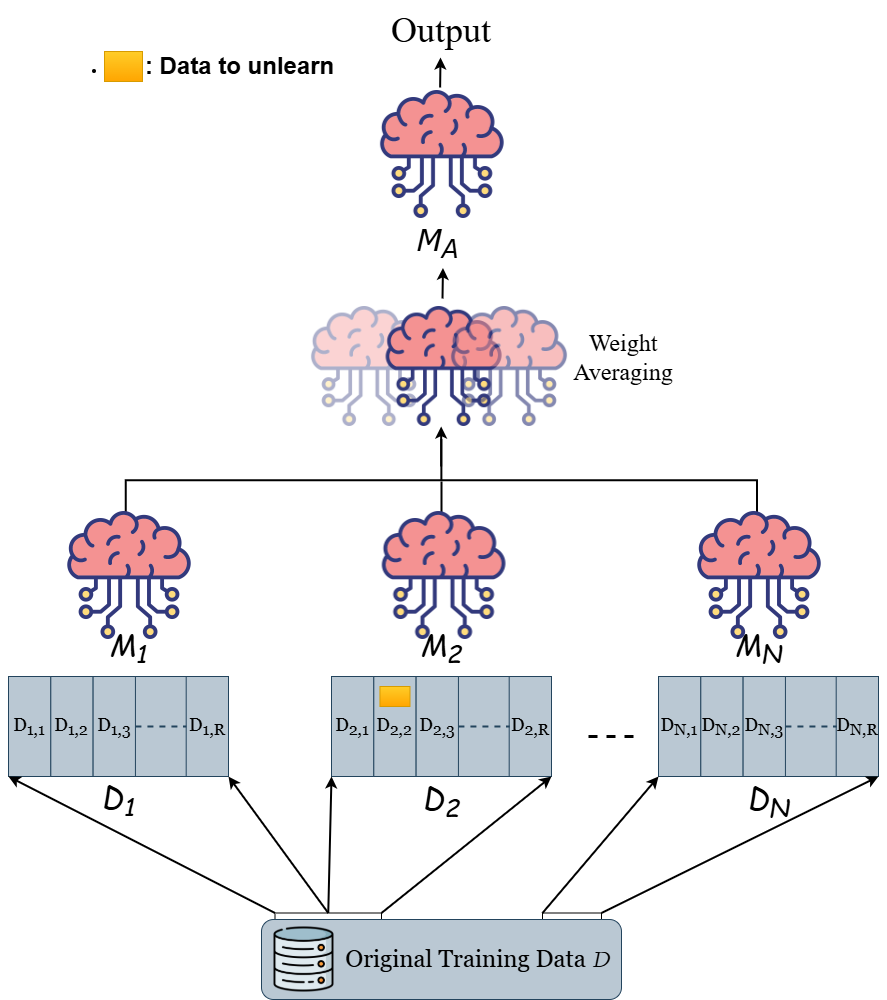}
     \caption{Workflow of \textbf{\texttt{SISA++}}; Original training dataset \( \mathcal{D} \) and its multiple shards (\( \mathcal{D}_1, \mathcal{D}_2, \dots, \mathcal{D}_N \)), which are further divided into slices (\( \mathcal{D}_{1,1}, \mathcal{D}_{1,2}, \dots, \mathcal{D}_{N,R} \)); Constituent model for each shard (\( M_1, M_2, \dots, M_N \)); Orange symbol indicates the data point to be unlearned; (\( M_A\)) represents the final model after weight averaging}
    \label{sisa++}
\end{figure}

\subsection{SISA++}
\textbf{\texttt{SISA++}} is a novel extension of SISA that keeps its structured partitioning strategy and follows the same steps as in SISA during the training time and as well as during retraining of models on certain shards.% while introducing a weight-averaging mechanism for model aggregation at inference time. 
\textbf{\texttt{SISA++}} algorithm is given in Algorithm \ref{alg:SISA++} and the workflow diagram in Figure \ref{sisa++}. %During training time, the dataset \( D = \{(x_i, y_i)\}_{i=1}^{N} \) is split into \( K \) independent disjoint shards \( D_k \), each with independent sub-models \( \mathcal{M}_i \) trained on the individual shard. Unlearning remains efficient since only the sub-models for impacted shards need to be retrained when a data point is deleted or modified. 
The critical point of differentiation in \texttt{\textbf{SISA++}} is inference: rather than majority vote or plain averaging, it combines knowledge through weight averaging \cite{wortsman2022model, 10095147}. The final model \( \mathcal{M}_{\text{A}} \) is obtained by averaging these accumulated model weights across all shards. This maintains performance without introducing any extra computational cost, such as longer inference time, ensuring stable and consistent predictions. 

\begin{algorithm}
\caption{\texttt{\textbf{SISA++}}}
\begin{algorithmic}[1]
\Require A set of models $\{\mathcal{M}_1, \mathcal{M}_2, \dots, \mathcal{M}_N\}$, each trained on a distinct data shard; $N$ is the total shards
\Ensure Final merged model $\mathcal{M}_{\text{A}}$
\State Initialize aggregated model: $\mathcal{M}_{\text{N}} \gets 0$
\For{each model $\mathcal{M}_i \in \{\mathcal{M}_1, \dots, \mathcal{M}_N\}$}
    \State $\mathcal{M}_{\text{N}} \gets \mathcal{M}_{\text{N}} + \mathcal{M}_i$
\EndFor
\State Compute weight averaging:
\State $\mathcal{M}_{\text{A}} \gets \frac{1}{N} \, \mathcal{M}_{\text{N}}$
\State \Return $\mathcal{M}_{\text{A}}$
\end{algorithmic}
\label{alg:SISA++}
\end{algorithm}

\begin{table*}[hbt!]
\setlength{\tabcolsep}{6pt}
\scriptsize
\centering
\begin{tabular}{l|cccccc|ccccccc}
\toprule
\multirow{2}{*}{Features} & \multicolumn{6}{c}{\textbf{E-DAIC}} & \multicolumn{6}{c}{\textbf{CREMA-D}} \\[0.5ex]
\cmidrule(lr){2-7} \cmidrule(lr){8-13}
 & \multicolumn{2}{c}{\textbf{SVM}} & \multicolumn{2}{c}{\textbf{CNN}} & \multicolumn{2}{c}{\textbf{TRA}} & \multicolumn{2}{c}{\textbf{SVM}} & \multicolumn{2}{c}{\textbf{CNN}} & \multicolumn{2}{c}{\textbf{TRA}} \\
\cmidrule(lr){2-3} \cmidrule(lr){4-5} \cmidrule(lr){6-7} \cmidrule(lr){8-9} \cmidrule(lr){10-11} \cmidrule(lr){12-13}
 & \textbf{M} \(\downarrow\) &  \textbf{R} \(\downarrow\) & \textbf{M} \(\downarrow\) &  \textbf{R} \(\downarrow\) & \textbf{M} \(\downarrow\) &  \textbf{R} \(\downarrow\) & \textbf{A(\%)} \(\uparrow\) & \textbf{F1 (\%)} \(\uparrow\) & \textbf{A(\%)} \(\uparrow\) & \textbf{F1 (\%)} \(\uparrow\) & \textbf{A(\%)} \(\uparrow\) & \textbf{F1(\%)} \(\uparrow\) \\
\midrule
\textbf{x-vector}  & 6.31 & 7.60  & 4.17 & 5.29  & 3.80 & 5.75  & 62.17 & 61.30 & 65.37 & 64.54 & 71.32 & 70.45 \\
\textbf{XLS-R}     & 6.35 & 7.41  & 4.28 & 5.10  & 3.92 & 4.47  & 64.23 & 63.89 & 69.31 & 67.83 & 76.83 & 75.99 \\
\textbf{TRILLsson} & 5.76 & 7.24  & 4.25 & 4.83  & \textbf{3.59} & \textbf{
4.43} & 67.82 & 66.14 & 68.52 & 67.35 & \textbf{77.69} & \textbf{76.02} \\
\textbf{WavLM}    & 7.60 & 7.34  & 4.25 & 5.70  & 3.76 & 4.66  & 63.78 & 62.91 & 68.84 & 67.61 & 76.75 & 74.29 \\
\textbf{MFCC}     & 7.96 & 8.23  & 4.39 & 6.08  & 4.74 & 6.10  & 39.67 & 37.55 & 44.24 & 43.19 & 51.69 & 48.64 \\
\bottomrule
\end{tabular}
\caption{Evalution Scores of different features with different downstream networks; TRA Stands for Transformer; M, R, A, F1 stands for MAE, RMSE, Accuracy, marco average F1 score; The abbreviations used in this Table are also used in Table \ref{tab:after}; The values in \textbf{{BOLD}} indicate the top model in that particular dataset}
\label{tab:before}
\end{table*}

\begin{table*}[hbt!]
\setlength{\tabcolsep}{7pt}
\scriptsize
\centering
\begin{tabular}{l|cc|cc|cc||cc|cc|cc}
    \toprule
    & \multicolumn{6}{c||}{\textbf{E-DAIC}} & \multicolumn{6}{c}{\textbf{CREMA-D}} \\
    \cmidrule(lr){2-7} \cmidrule(lr){8-13}
   & \multicolumn{2}{c}{\textbf{SVM}} & \multicolumn{2}{c}{\textbf{CNN}} & \multicolumn{2}{c}{\textbf{TRA}} & \multicolumn{2}{c}{\textbf{SVM}} & \multicolumn{2}{c}{\textbf{CNN}} & \multicolumn{2}{c}{\textbf{TRA}} \\
\cmidrule(lr){2-3} \cmidrule(lr){4-5} \cmidrule(lr){6-7} \cmidrule(lr){8-9} \cmidrule(lr){10-11} \cmidrule(lr){12-13}
 & \textbf{M} \(\downarrow\) &  \textbf{R} \(\downarrow\) & \textbf{M} \(\downarrow\) &  \textbf{R} \(\downarrow\) & \textbf{M} \(\downarrow\) &  \textbf{R} \(\downarrow\) & \textbf{A(\%)} \(\uparrow\) & \textbf{F1 (\%)} \(\uparrow\) & \textbf{A(\%)} \(\uparrow\) & \textbf{F1 (\%)} \(\uparrow\) & \textbf{A(\%)} \(\uparrow\) & \textbf{F1(\%)} \(\uparrow\) \\
\midrule
    % \cmidrule(lr){2-3} \cmidrule(lr){4-5} \cmidrule(lr){6-7} \cmidrule(lr){8-9} \cmidrule(lr){10-11} \cmidrule(lr){12-13}
    \multicolumn{13}{c}{\textbf{\textcolor{blue}{S I S A}}} \\
    \midrule
    \multicolumn{13}{c}{\textbf{1 USER REMOVED (4-shard)}} \\
    \midrule
    % --- SISA, 1 USER REMOVED (4-shard)
    \textbf{x-vector} & 
      \cellcolor{mustard10}6.92 & \cellcolor{mustard10}8.11 & 
      \cellcolor{mustard10}4.83 & \cellcolor{mustard10}6.20 & 
      \cellcolor{mustard10}4.32 & \cellcolor{mustard5}6.03 &
      \cellcolor{mustard15}61.34 & \cellcolor{mustard15}60.81 & 
      \cellcolor{mustard5}63.47 & \cellcolor{mustard5}62.18 & 
      \cellcolor{mustard5}68.25 & \cellcolor{mustard5}67.26 \\
    \textbf{XLS-R} &
      \cellcolor{mustard10}6.85 & \cellcolor{mustard10}7.74 & 
      \cellcolor{mustard10}4.81 & \cellcolor{mustard10}6.12 & 
      \cellcolor{mustard10}4.32 & \cellcolor{mustard10}5.58 &
      \cellcolor{mustard10}57.25 & \cellcolor{mustard10}56.47 & 
      \cellcolor{mustard10}64.78 & \cellcolor{mustard10}63.34 & 
      \cellcolor{mustard10}70.14 & \cellcolor{mustard10}69.08 \\
    \textbf{TRILLsson} &
      \cellcolor{mustard15}6.46 & \cellcolor{mustard15}7.63 & 
      \cellcolor{mustard15}4.70 & \cellcolor{mustard15}5.71 & 
      \cellcolor{mustard15}\textbf{4.23} & \cellcolor{mustard15}\textbf{5.52} &
      \cellcolor{mustard10}56.14 & \cellcolor{mustard10}55.81 & 
      \cellcolor{mustard15}65.33 & \cellcolor{mustard15}64.87 & 
      \cellcolor{mustard15}\textbf{72.36} & \cellcolor{mustard15}\textbf{71.58} \\
    \textbf{WavLM} &
      \cellcolor{mustard5}7.70 & \cellcolor{mustard5}8.41 & 
      \cellcolor{mustard5}5.54 & \cellcolor{mustard5}6.53 & 
      \cellcolor{mustard5}4.96 & \cellcolor{mustard10}5.73 &
      \cellcolor{mustard5}52.21 & \cellcolor{mustard5}51.34 & 
      \cellcolor{mustard10}65.21 & \cellcolor{mustard10}64.28 & 
      \cellcolor{mustard10}70.28 & \cellcolor{mustard10}69.16 \\
    \textbf{MFCC} &
      \cellcolor{mustard5}8.73 & \cellcolor{mustard5}9.91 & 
      \cellcolor{mustard5}5.55 & \cellcolor{mustard5}7.14 & 
      \cellcolor{mustard5}5.14 & \cellcolor{mustard5}6.83 &
      \cellcolor{mustard5}29.25 & \cellcolor{mustard5}28.89 & 
      \cellcolor{mustard5}41.60 & \cellcolor{mustard5}40.90 & 
      \cellcolor{mustard5}45.75 & \cellcolor{mustard5}44.28 \\
    \midrule
    \multicolumn{13}{c}{\textbf{2 USER REMOVED (4-shard)}} \\
    \midrule
    % --- SISA, 2 USER REMOVED (4-shard)
    \textbf{x-vector} &
      \cellcolor{mustard10}7.12 & \cellcolor{mustard10}8.43 & 
      \cellcolor{mustard10}4.93 & \cellcolor{mustard10}6.42 & 
      \cellcolor{mustard10}4.40 & \cellcolor{mustard10}6.25 &
      \cellcolor{mustard15}58.31 & \cellcolor{mustard15}57.26 & 
      \cellcolor{mustard5}63.07 & \cellcolor{mustard5}61.63 & 
      \cellcolor{mustard5}64.26 & \cellcolor{mustard5}63.64 \\
    \textbf{XLS-R} &
      \cellcolor{mustard10}7.08 & \cellcolor{mustard10}8.31 & 
      \cellcolor{mustard10}5.00 & \cellcolor{mustard10}6.53 & 
      \cellcolor{mustard15}4.53 & \cellcolor{mustard10}6.04 &
      \cellcolor{mustard10}57.25 & \cellcolor{mustard10}56.85 & 
      \cellcolor{mustard10}64.13 & \cellcolor{mustard10}63.22 & 
      \cellcolor{mustard15}70.25 & \cellcolor{mustard15}69.64 \\
    \textbf{TRILLsson} &
      \cellcolor{mustard15}6.74 & \cellcolor{mustard15}8.03 & 
      \cellcolor{mustard15}4.91 & \cellcolor{mustard15}6.32 & 
      \cellcolor{mustard15}\textbf{4.31} & \cellcolor{mustard15}\textbf{5.82} &
      \cellcolor{mustard15}56.85 & \cellcolor{mustard15}55.21 & 
      \cellcolor{mustard15}66.58 & \cellcolor{mustard15}65.97 & 
      \cellcolor{mustard15}\textbf{71.36} & \cellcolor{mustard15}\textbf{70.64} \\
    \textbf{WavLM} &
      \cellcolor{mustard5}7.92 & \cellcolor{mustard5}8.53 & 
      \cellcolor{mustard5}5.42 & \cellcolor{mustard5}6.73 & 
      \cellcolor{mustard5}5.02 & \cellcolor{mustard5}6.28 &
      \cellcolor{mustard5}52.36 & \cellcolor{mustard5}51.63 & 
      \cellcolor{mustard10}67.01 & \cellcolor{mustard10}65.92 & 
      \cellcolor{mustard5}70.63 & \cellcolor{mustard5}69.15 \\
    \textbf{MFCC} &
      \cellcolor{mustard5}8.87 & \cellcolor{mustard5}10.27 & 
      \cellcolor{mustard5}5.74 & \cellcolor{mustard5}7.35 & 
      \cellcolor{mustard5}5.32 & \cellcolor{mustard5}7.02 &
      \cellcolor{mustard5}27.28 & \cellcolor{mustard5}26.15 & 
      \cellcolor{mustard5}40.17 & \cellcolor{mustard5}40.39 & 
      \cellcolor{mustard5}42.63 & \cellcolor{mustard5}41.96 \\
    \midrule
    \multicolumn{13}{c}{\textbf{1 USER REMOVED (8-shard)}} \\
    \midrule
    % --- SISA, 1 USER REMOVED (8-shard)
    \textbf{x-vector} &
      \cellcolor{mustard10}6.56 & \cellcolor{mustard10}8.32 & 
      \cellcolor{mustard10}5.03 & \cellcolor{mustard10}7.23 & 
      \cellcolor{mustard10}4.54 & \cellcolor{mustard10}6.32 &
      \cellcolor{mustard5}56.22 & \cellcolor{mustard5}55.02 & 
      \cellcolor{mustard5}62.08 & \cellcolor{mustard5}59.36 & 
      \cellcolor{mustard5}61.81 & \cellcolor{mustard5}60.02 \\
    \textbf{XLS-R} &
      \cellcolor{mustard10}6.03 & \cellcolor{mustard10}8.10 & 
      \cellcolor{mustard10}4.94 & \cellcolor{mustard10}7.01 & 
      \cellcolor{mustard10}4.44 & \cellcolor{mustard10}6.00 &
      \cellcolor{mustard10}56.74 & \cellcolor{mustard10}55.37 & 
      \cellcolor{mustard10}63.88 & \cellcolor{mustard10}61.27 & 
      \cellcolor{mustard15}71.52 & \cellcolor{mustard15}70.61 \\
    \textbf{TRILLsson} &
      \cellcolor{mustard15}5.92 & \cellcolor{mustard15}7.82 & 
      \cellcolor{mustard15}4.80 & \cellcolor{mustard15}6.72 & 
      \cellcolor{mustard15}\textbf{4.35} & \cellcolor{mustard15}\textbf{5.92} &
      \cellcolor{mustard15}57.91 & \cellcolor{mustard15}56.63 & 
      \cellcolor{mustard15}66.24 & \cellcolor{mustard15}64.79 & 
      \cellcolor{mustard15}\textbf{72.80} & \cellcolor{mustard15}\textbf{71.53} \\
    \textbf{WavLM} &
      \cellcolor{mustard5}8.21 & \cellcolor{mustard5}10.53 & 
      \cellcolor{mustard5}7.12 & \cellcolor{mustard5}9.10 & 
      \cellcolor{mustard5}6.51 & \cellcolor{mustard5}8.05 &
      \cellcolor{mustard5}57.85 & \cellcolor{mustard5}56.28 & 
      \cellcolor{mustard5}65.49 & \cellcolor{mustard5}64.90 & 
      \cellcolor{mustard5}71.80 & \cellcolor{mustard5}70.63 \\
    \textbf{MFCC} &
      \cellcolor{mustard5}8.71 & \cellcolor{mustard5}10.92 & 
      \cellcolor{mustard5}7.49 & \cellcolor{mustard5}9.33 & 
      \cellcolor{mustard5}7.01 & \cellcolor{mustard5}8.83 &
      \cellcolor{mustard5}26.12 & \cellcolor{mustard5}25.85 & 
      \cellcolor{mustard5}39.63 & \cellcolor{mustard5}38.11 & 
      \cellcolor{mustard5}43.22 & \cellcolor{mustard5}42.61 \\
    \midrule
    \multicolumn{13}{c}{\textbf{2 USER REMOVED (8-shard)}} \\
    \midrule
    % --- SISA, 2 USER REMOVED (8-shard)
    \textbf{x-vector} &
      \cellcolor{mustard10}6.33 & \cellcolor{mustard10}8.54 & 
      \cellcolor{mustard10}5.22 & \cellcolor{mustard10}7.33 & 
      \cellcolor{mustard10}4.62 & \cellcolor{mustard10}6.44 &
      \cellcolor{mustard5}55.38 & \cellcolor{mustard5}54.17 & 
      \cellcolor{mustard5}61.87 & \cellcolor{mustard5}60.35 & 
      \cellcolor{mustard5}61.94 & \cellcolor{mustard5}60.98 \\
    \textbf{XLS-R} &
      \cellcolor{mustard10}6.21 & \cellcolor{mustard10}8.32 & 
      \cellcolor{mustard10}5.13 & \cellcolor{mustard10}7.11 & 
      \cellcolor{mustard10}4.51 & \cellcolor{mustard10}6.04 &
      \cellcolor{mustard10}56.61 & \cellcolor{mustard10}55.43 & 
      \cellcolor{mustard10}63.31 & \cellcolor{mustard10}62.60 & 
      \cellcolor{mustard15}70.85 & \cellcolor{mustard15}69.44 \\
    \textbf{TRILLsson} &
      \cellcolor{mustard15}6.12 & \cellcolor{mustard15}8.04 & 
      \cellcolor{mustard15}5.06 & \cellcolor{mustard15}7.04 & 
      \cellcolor{mustard15}\textbf{4.45} & \cellcolor{mustard15}\textbf{6.02} &
      \cellcolor{mustard15}55.38 & \cellcolor{mustard15}54.17 & 
      \cellcolor{mustard15}66.19 & \cellcolor{mustard15}64.72 & 
      \cellcolor{mustard15}\textbf{72.17} & \cellcolor{mustard15}\textbf{71.84} \\
    \textbf{WavLM} &
      \cellcolor{mustard5}8.43 & \cellcolor{mustard5}10.73 & 
      \cellcolor{mustard5}7.23 & \cellcolor{mustard5}9.23 & 
      \cellcolor{mustard5}6.65 & \cellcolor{mustard5}8.13 &
      \cellcolor{mustard5}55.26 & \cellcolor{mustard5}54.30 & 
      \cellcolor{mustard5}65.69 & \cellcolor{mustard5}63.76 & 
      \cellcolor{mustard5}68.34 & \cellcolor{mustard5}67.56 \\
    \textbf{MFCC} &
      \cellcolor{mustard5}8.84 & \cellcolor{mustard5}11.04 & 
      \cellcolor{mustard5}7.59 & \cellcolor{mustard5}9.53 & 
      \cellcolor{mustard5}7.00 & \cellcolor{mustard5}9.03 &
      \cellcolor{mustard5}26.01 & \cellcolor{mustard5}25.64 & 
      \cellcolor{mustard5}39.31 & \cellcolor{mustard5}38.01 & 
      \cellcolor{mustard5}42.48 & \cellcolor{mustard5}42.76 \\
    \midrule
    \multicolumn{13}{c}{\textbf{\textcolor{blue}{S I S A + +}}} \\
    \midrule
    \multicolumn{13}{c}{\textbf{1 USER REMOVED (4-shard)}} \\
    \midrule
    % --- SISA++ , 1 USER REMOVED (4-shard)
    \textbf{x-vector} &
      \cellcolor{mustardPP10}6.66 & \cellcolor{mustardPP10}7.90 & 
      \cellcolor{mustardPP10}4.60 & \cellcolor{mustardPP10}6.04 & 
      \cellcolor{mustardPP10}4.02 & \cellcolor{mustardPP10}5.73 &
      \cellcolor{mustardPP20}63.29 & \cellcolor{mustardPP20}62.52 & 
      \cellcolor{mustardPP5}64.22 & \cellcolor{mustardPP5}63.01 & 
      \cellcolor{mustardPP5}71.24 & \cellcolor{mustardPP5}70.61 \\
    \textbf{XLS-R} &
      \cellcolor{mustardPP20}6.12 & \cellcolor{mustardPP20}7.51 & 
      \cellcolor{mustardPP20}4.52 & \cellcolor{mustardPP20}5.83 & 
      \cellcolor{mustardPP20}3.96 & \cellcolor{mustardPP20}5.30 &
      \cellcolor{mustardPP15}61.24 & \cellcolor{mustardPP15}60.47 & 
      \cellcolor{mustardPP15}64.13 & \cellcolor{mustardPP15}63.22 & 
      \cellcolor{mustardPP15}70.25 & \cellcolor{mustardPP15}69.64 \\
    \textbf{TRILLsson} &
      \cellcolor{mustardPP15}6.44 & \cellcolor{mustardPP15}7.31 & 
      \cellcolor{mustardPP15}4.46 & \cellcolor{mustardPP15}5.41 & 
      \cellcolor{mustardPP20}\textbf{3.91} & \cellcolor{mustardPP20}\textbf{5.22} &
      \cellcolor{mustardPP10}60.88 & \cellcolor{mustardPP10}59.21 & 
      \cellcolor{mustardPP20}66.58 & \cellcolor{mustardPP20}65.97 & 
      \cellcolor{mustardPP20}\textbf{75.39} & \cellcolor{mustardPP20}\textbf{74.64} \\
    \textbf{WavLM} &
      \cellcolor{mustardPP10}7.47 & \cellcolor{mustardPP10}8.09 & 
      \cellcolor{mustardPP10}5.00 & \cellcolor{mustardPP10}6.31 & 
      \cellcolor{mustardPP10}4.59 & \cellcolor{mustardPP10}5.39 &
      \cellcolor{mustardPP10}52.36 & \cellcolor{mustardPP10}51.63 & 
      \cellcolor{mustardPP10}67.01 & \cellcolor{mustardPP10}65.92 & 
      \cellcolor{mustardPP10}70.63 & \cellcolor{mustardPP10}69.75 \\
    \textbf{MFCC} &
      \cellcolor{mustardPP5}8.32 & \cellcolor{mustardPP5}9.58 & 
      \cellcolor{mustardPP5}5.20 & \cellcolor{mustardPP5}6.88 & 
      \cellcolor{mustardPP5}4.92 & \cellcolor{mustardPP5}6.67 &
      \cellcolor{mustardPP5}29.28 & \cellcolor{mustardPP5}28.96 & 
      \cellcolor{mustardPP5}42.17 & \cellcolor{mustardPP5}41.40 & 
      \cellcolor{mustardPP5}47.63 & \cellcolor{mustardPP5}46.96 \\
    \midrule
    \multicolumn{13}{c}{\textbf{2 USER REMOVED (4-shard)}} \\
    \midrule
    % --- SISA++ , 2 USER REMOVED (4-shard)
    \textbf{x-vector} &
      \cellcolor{mustardPP10}6.87 & \cellcolor{mustardPP10}8.41 & 
      \cellcolor{mustardPP10}4.79 & \cellcolor{mustardPP10}6.25 & 
      \cellcolor{mustardPP10}4.16 & \cellcolor{mustardPP10}5.97 &
      \cellcolor{mustardPP20}63.69 & \cellcolor{mustardPP20}62.85 & 
      \cellcolor{mustardPP5}64.83 & \cellcolor{mustardPP5}63.28 & 
      \cellcolor{mustardPP5}69.34 & \cellcolor{mustardPP5}68.64 \\
    \textbf{XLS-R} &
      \cellcolor{mustardPP15}6.60 & \cellcolor{mustardPP15}7.69 & 
      \cellcolor{mustardPP15}4.65 & \cellcolor{mustardPP15}5.89 & 
      \cellcolor{mustardPP15}4.09 & \cellcolor{mustardPP15}5.48 &
      \cellcolor{mustardPP10}60.79 & \cellcolor{mustardPP10}59.64 & 
      \cellcolor{mustardPP10}64.99 & \cellcolor{mustardPP10}64.02 & 
      \cellcolor{mustardPP20}73.85 & \cellcolor{mustardPP20}72.96 \\
    \textbf{TRILLsson} &
      \cellcolor{mustardPP20}6.32 & \cellcolor{mustardPP20}7.44 & 
      \cellcolor{mustardPP20}4.59 & \cellcolor{mustardPP20}5.40 & 
      \cellcolor{mustardPP20}\textbf{4.05} & \cellcolor{mustardPP20}\textbf{5.34} &
      \cellcolor{mustardPP15}62.74 & \cellcolor{mustardPP15}61.35 & 
      \cellcolor{mustardPP20}67.24 & \cellcolor{mustardPP20}66.41 & 
      \cellcolor{mustardPP20}\textbf{73.88} & \cellcolor{mustardPP20}\textbf{73.75} \\
    \textbf{WavLM} &
      \cellcolor{mustardPP10}7.54 & \cellcolor{mustardPP10}8.36 & 
      \cellcolor{mustardPP10}5.13 & \cellcolor{mustardPP10}6.41 & 
      \cellcolor{mustardPP10}4.85 & \cellcolor{mustardPP10}5.53 &
      \cellcolor{mustardPP10}59.64 & \cellcolor{mustardPP10}58.14 & 
      \cellcolor{mustardPP10}67.59 & \cellcolor{mustardPP10}66.80 & 
      \cellcolor{mustardPP10}74.14 & \cellcolor{mustardPP10}73.92 \\
    \textbf{MFCC} &
      \cellcolor{mustardPP5}8.44 & \cellcolor{mustardPP5}9.87 & 
      \cellcolor{mustardPP5}5.35 & \cellcolor{mustardPP5}7.05 & 
      \cellcolor{mustardPP5}5.07 & \cellcolor{mustardPP5}6.83 &
      \cellcolor{mustardPP5}33.96 & \cellcolor{mustardPP5}32.84 & 
      \cellcolor{mustardPP5}43.19 & \cellcolor{mustardPP5}42.05 & 
      \cellcolor{mustardPP5}46.24 & \cellcolor{mustardPP5}45.41 \\
    \midrule
    \multicolumn{13}{c}{\textbf{1 USER REMOVED (8-shard)}} \\
    \midrule
    % --- SISA++ , 1 USER REMOVED (8-shard)
    \textbf{x-vector} &
      \cellcolor{mustardPP10}6.53 & \cellcolor{mustardPP10}8.20 & 
      \cellcolor{mustardPP10}4.90 & \cellcolor{mustardPP10}7.11 & 
      \cellcolor{mustardPP10}4.42 & \cellcolor{mustardPP10}6.23 &
      \cellcolor{mustardPP10}59.03 & \cellcolor{mustardPP10}58.05 & 
      \cellcolor{mustardPP10}62.61 & \cellcolor{mustardPP10}62.61 & 
      \cellcolor{mustardPP10}64.08 & \cellcolor{mustardPP10}61.61 \\
    \textbf{XLS-R} &
      \cellcolor{mustardPP15}5.93 & \cellcolor{mustardPP15}8.03 & 
      \cellcolor{mustardPP15}4.82 & \cellcolor{mustardPP15}6.81 & 
      \cellcolor{mustardPP15}4.34 & \cellcolor{mustardPP15}6.12 &
      \cellcolor{mustardPP15}58.41 & \cellcolor{mustardPP15}59.74 & 
      \cellcolor{mustardPP15}64.74 & \cellcolor{mustardPP15}62.96 & 
      \cellcolor{mustardPP20}73.85 & \cellcolor{mustardPP20}72.64 \\
    \textbf{TRILLsson} &
      \cellcolor{mustardPP20}5.82 & \cellcolor{mustardPP20}6.93 & 
      \cellcolor{mustardPP20}4.73 & \cellcolor{mustardPP20}6.22 & 
      \cellcolor{mustardPP20}\textbf{4.24} & \cellcolor{mustardPP20}\textbf{5.73} &
      \cellcolor{mustardPP20}60.62 & \cellcolor{mustardPP20}57.64 & 
      \cellcolor{mustardPP20}66.41 & \cellcolor{mustardPP20}65.28 & 
      \cellcolor{mustardPP20}\textbf{74.96} & \cellcolor{mustardPP20}\textbf{73.21} \\
    \textbf{WavLM} &
      \cellcolor{mustardPP10}8.11 & \cellcolor{mustardPP10}10.47 & 
      \cellcolor{mustardPP10}7.00 & \cellcolor{mustardPP10}9.03 & 
      \cellcolor{mustardPP10}6.42 & \cellcolor{mustardPP10}7.83 &
      \cellcolor{mustardPP10}58.94 & \cellcolor{mustardPP10}57.37 & 
      \cellcolor{mustardPP10}66.73 & \cellcolor{mustardPP10}65.34 & 
      \cellcolor{mustardPP10}73.14 & \cellcolor{mustardPP10}72.87 \\
    \textbf{MFCC} &
      \cellcolor{mustardPP5}8.44 & \cellcolor{mustardPP5}10.77 & 
      \cellcolor{mustardPP5}7.22 & \cellcolor{mustardPP5}9.13 & 
      \cellcolor{mustardPP5}6.70 & \cellcolor{mustardPP5}8.11 &
      \cellcolor{mustardPP5}34.47 & \cellcolor{mustardPP5}33.61 & 
      \cellcolor{mustardPP5}42.36 & \cellcolor{mustardPP5}40.74 & 
      \cellcolor{mustardPP5}45.31 & \cellcolor{mustardPP5}44.47 \\
    \midrule
    \multicolumn{13}{c}{\textbf{2 USER REMOVED (8-shard)}} \\
    \midrule
    % --- SISA++ , 2 USER REMOVED (8-shard)
    \textbf{x-vector} &
      \cellcolor{mustardPP10}6.20 & \cellcolor{mustardPP10}8.44 & 
      \cellcolor{mustardPP10}5.11 & \cellcolor{mustardPP10}7.23 & 
      \cellcolor{mustardPP10}4.51 & \cellcolor{mustardPP10}6.55 &
      \cellcolor{mustardPP10}58.46 & \cellcolor{mustardPP10}57.55 & 
      \cellcolor{mustardPP10}63.28 & \cellcolor{mustardPP10}62.17 & 
      \cellcolor{mustardPP10}62.78 & \cellcolor{mustardPP10}61.85 \\
    \textbf{XLS-R} &
      \cellcolor{mustardPP15}6.11 & \cellcolor{mustardPP15}8.22 & 
      \cellcolor{mustardPP15}5.00 & \cellcolor{mustardPP15}7.02 & 
      \cellcolor{mustardPP15}4.43 & \cellcolor{mustardPP15}6.02 &
      \cellcolor{mustardPP15}57.65 & \cellcolor{mustardPP15}56.43 & 
      \cellcolor{mustardPP15}64.73 & \cellcolor{mustardPP15}63.09 & 
      \cellcolor{mustardPP20}71.85 & \cellcolor{mustardPP20}70.64 \\
    \textbf{TRILLsson} &
      \cellcolor{mustardPP20}6.05 & \cellcolor{mustardPP20}7.02 & 
      \cellcolor{mustardPP20}4.90 & \cellcolor{mustardPP20}6.25 & 
      \cellcolor{mustardPP20}\textbf{4.30} & \cellcolor{mustardPP20}\textbf{5.82} &
      \cellcolor{mustardPP20}58.64 & \cellcolor{mustardPP20}57.85 & 
      \cellcolor{mustardPP20}67.36 & \cellcolor{mustardPP20}65.78 & 
      \cellcolor{mustardPP20}\textbf{73.51} & \cellcolor{mustardPP20}\textbf{72.47} \\
    \textbf{WavLM} &
      \cellcolor{mustardPP10}8.36 & \cellcolor{mustardPP10}10.67 & 
      \cellcolor{mustardPP10}7.12 & \cellcolor{mustardPP10}9.03 & 
      \cellcolor{mustardPP10}6.54 & \cellcolor{mustardPP10}8.01 &
      \cellcolor{mustardPP10}56.82 & \cellcolor{mustardPP10}55.33 & 
      \cellcolor{mustardPP10}66.61 & \cellcolor{mustardPP10}64.88 & 
      \cellcolor{mustardPP10}70.34 & \cellcolor{mustardPP10}69.14 \\
    \textbf{MFCC} &
      \cellcolor{mustardPP5}8.59 & \cellcolor{mustardPP5}10.97 & 
      \cellcolor{mustardPP5}7.43 & \cellcolor{mustardPP5}9.23 & 
      \cellcolor{mustardPP5}6.81 & \cellcolor{mustardPP5}8.34 &
      \cellcolor{mustardPP5}32.74 & \cellcolor{mustardPP5}31.64 & 
      \cellcolor{mustardPP5}41.04 & \cellcolor{mustardPP5}39.67 & 
      \cellcolor{mustardPP5}43.98 & \cellcolor{mustardPP5}42.98 \\
    \bottomrule
\end{tabular}
\caption{Performance comparison on the E-DAIC and CREMA-D datasets; For EDAIC, metrics are Mean Absolute Error (M) and Root Mean Squared Error (R); For CREMA-D, metrics are Accuracy (A) and marco average F1-score; Results are reported under different conditions: one or two users datapoints removed, with evaluations on four-shards and eight-shards, and for both SISA and \texttt{\textbf{SISA++}} settings}
\label{tab:after}
\end{table*}
\subsection{Cookbook Recipes}

%To facilitate the widespread adoption of MU in PSP applications, we provide insights in the form of systematic \textit{``cookbook recipes''}. 
These \textit{``cookbook recipes''} will serve as structured guidelines for researchers, covering the selection of effective feature representations and the design of downstream architectures. By incorporating these elements, our study acts as a resourceful guide for future advancements in the field. Below, we detail the specific feature sets and downstream model architectures used in our experiments, providing a basis for selecting the most suitable configurations.

\noindent\textbf{Feature Set}:  We use TRILLsson\footnote{\url{https://tfhub.dev/google/nonsemantic-speech-benchmark/trillsson4/1}} \cite{shor2022trillsson}, distilled from Conformer (CAP12) \cite{shor2022universal} and trained on AudioSet and Libri-light datasets. It excels in NOSS benchmark tasks such as SER, speaker recognition, synthetic speech detection. We consider XLS-R \cite{babu22_interspeech}, a self-supervised model pre-trained on 436k hours of multilingual speech data. We use 0.3B parameter variant\footnote{\url{https://huggingface.co/facebook/wav2vec2-xls-r-300m}}. Further, we include WavLM\footnote{\url{https://huggingface.co/microsoft/wavlm-base}} \cite{chen2022wavlm}, a SOTA model in the SUPERB benchmark across various speech tasks including paralinguistic providing. We use its base version with 94.70M parameters. Lastly, we consider x-vector\footnote{\url{https://huggingface.co/speechbrain/spkrec-xvect-voxceleb}} \cite{8461375}, originally designed for speaker recognition, has been effectively applied to SER \cite{9054317} and DD \cite{9746068}. For all models, the final hidden states are extracted and averaged pooled to obtain fixed-length feature vectors: 1024 for TRILLsson, 768 for WavLM, 1280 for XLS-R, 512 for x-vector. All the audio data is sampled at 16kHz before passing through these models for feature extraction. We also make use of MFCC features in our experiments.

\noindent \textbf{Downstream Modeling}: We use SVM, CNN, and Transformer as downstream modeling networks. For SVM, we used a linear SVM and kept the default parameters as given in the \textit{Scikit-learn} library. The CNN architecture consists of 1D convolutional layers with 64 and 128 filters (kernel size of 3) and maxpooling after consecutive 1D convlutional layer. The output is flattened, followed by a dense layer with 128 neurons and ending with a output layer with softmax or linear for classification or regression. The FCN flattens input features before passing through dense layers with 64 and 128 neurons. For transformer, we use the vanilla transformer encoder \cite{vaswani2017attention} with a single head and then flattening the features to be fed to a dense layer with 128 neurons and lastly, followed by a output layer. FCN models trainable parameters ranges from 0.8M to 1.5M followed by CNN with 1.1M to 1.7M depending on the input feature shape. %For model unlearning (MU), we experiment with baselines such as retraining and fine-tuning\footnote{\url{https://github.com/tamlhp/awesome-machine-unlearning}}. All models are trained using the Adam optimizer and categorical cross-entropy loss for efficient optimization.  

\section{Experiments}

\subsection{Dataset}
\noindent \textbf{Crowd-sourced Emotional Multimodal Actors Dataset (CREMA-D)} \cite{cao2014crema} is widely used SER dataset containing 7,442 utterances from 91 speakers (48 male, 43 female) across six emotions: Anger, Happiness, Sadness, Fear, Disgust, and Neutral. We use 80:20 split ratio for training and testing our models. \newline
\noindent\textbf{E-DAIC} \cite{ringeval2018avec} is a benchmark DD dataset comprising 275 clinical interview sessions (163 training, 56 development, 56 test) with a balanced gender distribution. For our experiments, we segment each session into 5-second audio clips. We use the official split for training and evaluating our models. \newline
\noindent \textbf{Training Details}: During initial training, we train the models for 20 epochs with learning rate 1e-3 and Adam as the optimizer with batch size of 32. We also use dropout and early stopping for preventing overfitting. After removal of data points from certain shards, we retrain the models associated with these shards with the same training details as mentioned above during initial training. 

\subsection{Analysis and Results}
% \vspace{-0.10cm}
\noindent \textbf{Before Unlearning Request}: We evaluate the baseline performance of models before unlearning to establish a reference for comparison. From Table \ref{tab:before}, we observe that TRILLsson with Transformer downstream attains the best performance in both E-DAIC and CREMA-D.  \newline
\noindent \textbf{After Unlearning Request}: We evaluate the efficacy of \texttt{\textbf{SISA++}} against SISA by simulating user removal under different sharding configurations, as summarized in Table \ref{tab:after}. Suppose, one or two users presents their request for removal of their corresponding datapoints from the training points. So, we simulate and analyze the impact of removing (i) one user datapoints from a single shard in both 4-shard and 8-shard setups and (ii) two users points from two different shards in the same configurations. In these setups, the dataset is divided into either 4 or 8 shards for assessing of the granularity of sharding on unlearning performance. We conduct experiments using various feature sets and downstream architectures for both SER and DD. For experiments with SISA, our results indicates that finer-grained sharding (i.e., 8 shards) mitigates sometimes performance degradation compared to 4-shards and attains comparable or better performance than 4-shards. This behavior can be observed across both one user and two user datapoints removal for SER and DD. This occurs because each shard contains fewer users, thereby reducing the overall impact of data removal. In comparing the results of SISA baseline with our proposed novel method, it can be observed that \texttt{\textbf{SISA++}} consistently outperforms SISA, demonstrating superior retention of performance post-unlearning. Moreover, TRILLsson exhibit higher robustness to unlearning compared to traditional features such MFCC and other neural features followed by Transformer as the best downstream network. We can see that TRILLsson with transformer downstream as the best performing pair for all the sharding and user datapoints removal settings. This is observed for experiments with both SISA and \textbf{\texttt{SISA++}} with TRILLsson and Transformer with \textbf{\texttt{SISA++}} as the best combination. This trend is evident in both SER and DD tasks. Overall, we can see that the performance generally becomes low after unlearning in comparison to before unlearning request (See Table \ref{tab:before}). However, \textbf{\texttt{SISA++}} we were able to match the performance even after data removal in a better manner than SISA. These findings highlight that SISA++ provides a more resilient unlearning framework for PSP. Also, \textbf{\texttt{SISA++}} maintains the same training time and achieves comparable or even lower inference time than SISA, all while offering significantly better performance retention post-unlearning. This efficiency, combined with its superior ability to preserve model performance, underscores the advantage of \textbf{\texttt{SISA++}} for MU in PSP. These advantages calls for the usage of textbf{\texttt{SISA++}} in diverse applications. % It maintains good performance even after the unlearning process, demonstrating its robustness and suitability for privacy-sensitive applications. These results pave the way for the application of SISA++ in domains where data privacy and efficient unlearning are critical requirements.

\section{Conclusion}
In this study, we pioneered MU for PSP and proposed \textbf{\texttt{SISA++}}, an novel extension to previous SOTA MU method SISA. It merges models trained on different shards via weight averaging, outperforming SISA in preserving performance post-unlearning. To aid future research, we provide \textit{``cookbook recipes''} and as a part of this we recommend the use of transformer-based downstream with TRILLsson features for MU in PSP as they offer the best performance retention and validated through our experiments. Our findings will facilitate the easier adoption of MU in PSP applications. Our research calls for exploration of MU to various speech processing applications where privacy is a concern. Our work will also act as reference as for speech processing domain as MU in relatively underexplored in speech processing compared to other domains such as vision and NLP. %As MU remains underexplored in speech compared to computer vision and NLP, our work serves as a reference for future research in the field.

\bibliographystyle{IEEEtran}
\bibliography{main}

\end{document}